\documentclass[12pt]{article}
\usepackage{epsf}
\usepackage[dvips]{graphicx}
\begin{document}
\begin{center}
{\bf\large{Exotic narrow resonance searches in the systems $K_s^0
p$, $K_s^0 \Lambda$ and $\Lambda p$ in pA-interactions at 10 GeV/c
 }}

\vskip 2mm P.Zh.Aslanyan$^{1,2 \dag}$

\vskip 5mm {\small (1) {\it Joint Institute for Nuclear Research }
\\
(2) {\it Yerevan State University }
\\
$\dag$ {\it E-mail: paslanian@jinr.ru }}

\end{center}

\begin{abstract}
  Experimental data from the 2m propane bubble chamber have been
analyzed to search for an exotic baryon states, in the $K_s^0 p$,
$K_s^0 \Lambda$ and $\Lambda p$ decay mode for the reaction
p+$C_3H_8$ at 10 GeV/c.

The invariant mass spectrum $\Lambda K^0_s$ observe a narrow peaks
at 1750, 1795,1850 MeV/$c^2$. The statistical significance of these
peaks has been estimated as 5.6, 3.3 and 3.0 S.D., respectively.
There are the small enhancements in  mass regions of (1650-1675) and
(1925-1950) MeV/c$^2$. These would be  candidates for the $N^0$ or
the $\Xi^0$ pentaquark states.

The $pK^0_s$ invariant mass spectrum shows resonant structures with
$M_{K_s^0 p}$=1540, 1613, 1821 MeV/$c^2$. The statistical
significance of these peaks have been estimated as 5.5,4.8 and 5.0
s.d., respectively. There are also small peaks in 1487(3 s.d.),1690(
3.6 s.d.), 1750(2.3 s.d.) and 1980(3.0 s.d.) MeV/$c^2$ mass regions.

The invariant mass spectrum S=-1 $\Lambda p$ observe a narrow peaks
at 2100, 2175,2285 and 2353 MeV/$c^2$. Their excess above background
by the second method is 6.9, 4.9, 3.8 and 2.9 S.D., respectively.
There is also a small peak in 2225( 2.2 s.d.) MeV/$c^2$ mass region.

 The investigation has been performed at the
Veksler and Baldin Laboratory of High Energies, JINR.
\end{abstract}


\section{introduction}
 Multi-quark states, glueballs and hybrids have been searched for experimentally
for a very long time, but none is established. Several models
(\cite{1}-\cite{5}) predict the multiplet structure and
characteristics of multi quark hadrons and pentaquarks for example
the chiral soliton model, the uncorrelated quark model, correlated
quark models, QCD sum rules, thermal models, lattice QCD  .

 Results from a wide range of recent
experiments\cite{6,7} are consistent with the  existence of an
exotic S=+1 resonance, the $\Theta+(1540)$ with a narrow width and a
mass near 1540 MeV cite{1}.

Preliminary results on a search for the $\Xi^0$ I=1/2 as  well as
for the $N^0$ or the $\Xi^0$ pentaquark states in the decay mode
$\Lambda K^0_s$ with the mass $1734\pm0.5\pm5$ MeV/$c^2$  is
presented in the article\cite{8}.  A narrow resonance significant
signal for $\Xi^0(1750)\to \Xi^- \pi^+$ and $N^0(1680)\to N \pi$
were observed in \cite{9,10}.

 Metastable strange  dibaryons were searched a long time ago at LHE JINR, too.
The effective mass spectra of 17 strange multiquark systems were
studied. Our group succeeded in finding resonance-like peaks
\cite{11,12} only in five of them $\Lambda p$, $\Lambda p \pi$,
$\Lambda \Lambda$, $\Lambda \Lambda p $, $\Lambda \pi^+ \pi^+$.

\section{Experiment}

The JINR 2m bubble chamber is the most suitable instrument for this
purpose \cite{7,13}. The experimental information of more than
700000 stereo photographs are used to select the events with $V^0$
strange particles. The effective mass distribution of 8657-events
with $\Lambda$, 4122-events with $K^0_s$ particles are consistent
with their PDG values \cite{7,13}.  The effective mass resolution of
$\Lambda K^0_s$ system  was estimated to be on the average 1\%. The
effective mass resolution for systems from $ K^0_s p$ and $\Lambda
p$ combinations were estimated to be 0.6 \% for protons over the
following momentum range: $0.150\le p\le $ 0.900 GeV/c.

\section{$pK^0_s$,$\Lambda K^0_s$ and $\Lambda p$ spectrum  analysis}

The total experimental background has been  obtained by three
methods \cite{7,13,14}. In the first method, the experimental
effective mass distribution was approximated by the polynomial
function  after cutting out the resonance ranges because this
procedure has to provide the fit with $\Xi^2$=1 and polynomial
coefficient with errors less than 30 \%.  The second of the randomly
mixing method of the angle between of  decaying particles from the
resonance for experimental events is described in \cite{17}. Then,
these background events were analyzed by using the same experimental
condition.  The third background method has been obtained  by using
FRITIOF model (\cite{18}) with experimental conditions. The analysis
done by three methods has shown that while fitting these
distributions had the same coefficients and order of polynomial. The
values for the mean position of the peak and the width obtained by
using Breit Wigner fits.

\subsection{$pK^0_s$ - spectrum for protons with a momentum of
$0.350\le p_p\le 0.900$ GeV/c }

  The $pK^0_s$ effective mass distribution 2300 combination (Fig.1a) is
shown resonant structures with $M_{K_s^0 p}$=1540, 1613, 1821
MeV/$c^2$ and $\Gamma_{K_s^0 p}$= 9.2, 16.1, 28.0
MeV/$c^2$(\cite{7}). The statistical significance of these peaks
have been estimated as 5.5,4.8 and 5.0 s.d., respectively. There are
also small peaks in 1690( 3.6 s.d.), 1750 (2.3 s.d.) and 1980(3.0
s.d.) MeV/$c^2$ mass regions.

\subsection{$\Lambda K^0_s$ - spectrum  analysis }

Figure 1b  shows  the invariant mass of 1012 ($\Lambda
K^0_s$)combinations with bin sizes 10 MeV/$c^2 $(\cite{13}).  There
are significant enhancements in mass regions of 1750, 1795 and 1850
MeV/$c^2$(Fig.1b). Their excess above background by the first method
is 5.0, 2.7 , 3.0 S.D.. There are small enhancement in mass regions
of 1670 and 1935 MeV/$c^2$.

\subsection{$\Lambda p$ - spectrum  analysis for protons with a momentum of
$0.250\le p_p\le 0.900$ GeV/c }

Figure  1c  shows  the invariant mass of 2434 ($\Lambda
p$)combinations with bin sizes 15 MeV/$c^2 $(\cite{12}). The values
for the mean position of the peak and the width obtained by using
Breit Wigner fits. There are significant enhancements in mass
regions of 2100, 2175, 2285 and 2353 MeV/$c^2$(Fig.1c).Their excess
above background by the second method is 6.9, 4.9, 3.8 and 2.9 S.D.,
respectively. There is also a small peak in 2225( 2.2 s.d.)
MeV/$c^2$ mass region.

\section{Conclusion}

A number of peculiarities were found in the effective mass spectrum
of: $K^0_s p$ in regions of 1487, 1540, 1685, 1750, 1821 and 1980
MeV/$c^2$: $\Lambda K^0_s$ in regions of 1670,1750, 1785,1850 and
1935 MeV/$c^2$; $\Lambda p$ in regions of 2100, 2175,2225,2285 and
2353 MeV/$c^2$.

These peaks in the effective mass spectrum $\Lambda K^0_s$ are
possible candidates for two pentaquark states: the N0 with quark
content udsds decaying into $\Lambda K^0$ and the $\Xi^0$  quark
content udssd decaying into $\Lambda \overline{K^0}$, which are
agreed: with the calculated  rotational spectra   N0  and $\Xi^0$
spectra from the theoretical report  of  D. Akers \cite{5},A.A.
Arkhipov \cite{16} and with $\Theta+$ spectra from  the experimental
reports of Yu.A.Troyan \cite{15} and P. Aslanyan \cite{7}.

 The experimental result for S=-1  $\Lambda p$  dibaryon spectrum shows
that the predicted peaks with the bag model has been
confirmed\cite{12}.

\begin{figure}
 \epsfysize=180mm
 \centerline{
 \epsfbox{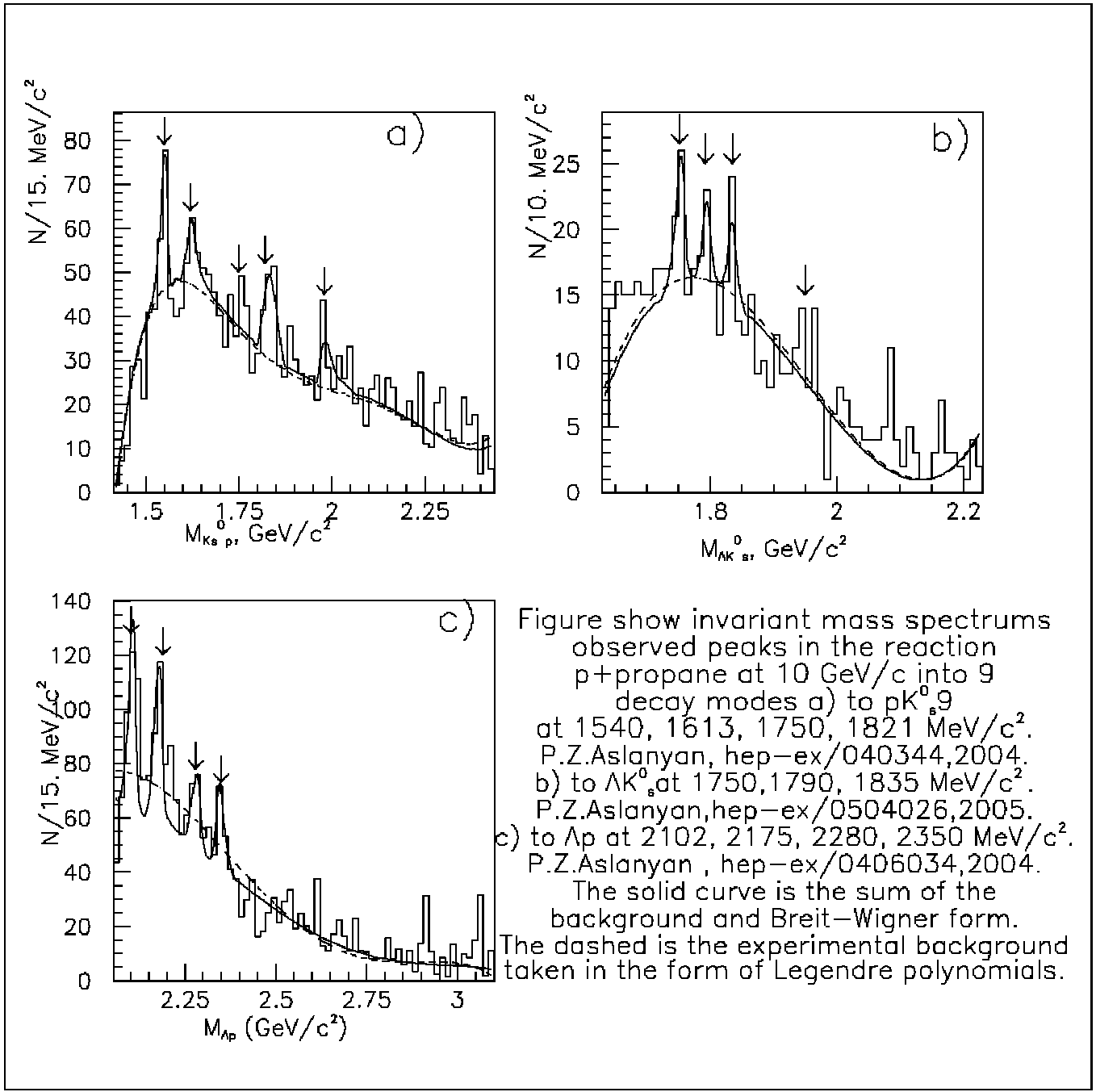}}
 \caption{The effective mass distribution for systems:(a) $pK^0_s$,
  (b)$\Lambda K^0_s$ and (c)$\Lambda p$. }
   \end{figure}


\begin{thebibliography}{9}
\bibitem{1} D. Diakonov, V. Petrov, and M. Polyakov, Z. Phys. A 359 , 305(1997).
\bibitem{2}V.Guzey and M.Polyakov, arXiv hep-ph/0501010,2005.
\bibitem{3} R. L. Jaffe, F. Wilczek, Phys. Rev. Lett. 91 (2003) 232003, hep-ph/0307341.
\bibitem{4}J. Ellis et al, JHEP 0405:002, 2004, hep-ph/0401127.
\bibitem{5} D. Akers, arXiv.org:hep-ph/0311031, 2004.
\bibitem{6}T.Nakano et al. [LEPS Collaboration], Phys.Rev.Lett.91(2003)012002, hep-ex/0301020.
\bibitem{7}P.Z.Aslanyan et al., hep-ex/0403044, 2004; JINR  Communications,
E1-2004-137,2004.
\bibitem{8}S.Kabana, Nuclear Dynamics 20th Winter
Workshop on Nuclear Dynamics Trelawny Beach, Jamaica March 15–20,
2004; hep-ex/0406032,2004.
\bibitem{9}ZEUS Collaboration, S.Chekanov et al.,hep-ex/0501069,2005.
\bibitem{10}GRAAL collab., hep-ex/0409032,2004
\bibitem{11}B.A. Shahbazian et al., Nucl. Physics, A374(1982),p. 73c-93.c.2
\bibitem{12}P.Z. Aslanyan et al., International Conference: I.Ya.Pomeranchuk and
Physics at the Turn of Centuries, Moscow, Russia, 24-28 January,
2003;e-Print Archive:hep-ex/0406034 .
\bibitem{13}P.Z.Aslanyan et al., hep-ex/0403044, 2005.
\bibitem{14}P.Z. Aslanian,et al.,hep-ex/0504026,2005.P.Z. Aslanian,et al., JINR
Commun. E1-2001-265,2001.
\bibitem{15}Yu.Troyan et. al.,JINR, D1-2004-39, Dubna,2004;hep-ex/0404003(2004).
\bibitem{16} A.A.Arkhipov: Archive:hep-ph/0403284v3,2004.
\bibitem{17} V.L.Lyuboshits at al., JINR Rapid Comm., N6(74),p209, 1995.
\bibitem{18} FRITIOF, H. Pi, Comput. Phys.Commun. 71,173, 1992.
A.S.Galoian et al., JINR Commun., P1-2002-54, 2002.
\end{thebibliography}
\end{document}